\documentclass[aps,prd,superscriptaddress,10pt,nofootinbib,twocolumn,preprintnumbers]{revtex4} 

\usepackage{amsfonts}
\usepackage{amssymb}
\usepackage{amsmath}
\usepackage{graphicx}
\usepackage{subcaption}
\begin{document}

\preprint{YITP-17-100, IPMU17-0126}

\title{Solving the flatness problem with an anisotropic instanton in Ho\v{r}ava-Lifshitz gravity}

\author{Sebastian F. Bramberger}
\affiliation{Max Planck Institute for Gravitational Physics (Albert Einstein Institute), 14476 Potsdam-Golm, Germany}
\author{Andrew Coates}
\affiliation{School of Mathematical Sciences, University of Nottingham, University Park, Nottingham, NG7 2RD, United Kingdom}
\author{Jo\~{a}o Magueijo}
\affiliation{Theoretical Physics, Blackett Laboratory, Imperial College, London, SW7 2BZ, United Kingdom}
\author{Shinji Mukohyama}
\affiliation{Center for Gravitational Physics, Yukawa Institute for Theoretical Physics, Kyoto University, 606-8502, Kyoto, Japan}
\affiliation{Kavli Institute for the Physics and Mathematics of the Universe (WPI), The University of Tokyo Institutes for Advanced Study, The University of Tokyo, Kashiwa, Chiba 277-8583, Japan}
\author{Ryo Namba}
\affiliation{Department of Physics, McGill University, Montr\'{e}al, QC, H3A 2T8, Canada}
\author{Yota Watanabe}
\affiliation{Kavli Institute for the Physics and Mathematics of the Universe (WPI), The University of Tokyo Institutes for Advanced Study, The University of Tokyo, Kashiwa, Chiba 277-8583, Japan}
\affiliation{Center for Gravitational Physics, Yukawa Institute for Theoretical Physics, Kyoto University, 606-8502, Kyoto, Japan}

\date{\today}

 \begin{abstract}
In Ho\v{r}ava-Lifshitz gravity a scaling isotropic in space but anisotropic in spacetime, often called anisotropic scaling with the dynamical critical exponent $z=3$, lies at the base of its renormalizability. This scaling also leads to a novel mechanism of generating scale-invariant cosmological perturbations, solving the horizon problem without inflation. In this paper we propose a possible solution to the flatness problem, in which we assume that the initial condition of the Universe is set by a small instanton respecting the same scaling. We argue that the mechanism may be more general than the concrete model presented here, and rely simply on the deformed dispersion relations of the theory, and on equipartition of the various forms of energy at the starting point. 
 \end{abstract}

\maketitle

\section{Introduction}\label{intro}

In general relativity a homogeneous and isotropic universe is described by the Friedmann equation
\begin{equation}\label{fried0}
 3 H^2 = 8\pi G \rho - \frac{3K}{a^2} + \Lambda\,,
\end{equation}
where $H$ is the Hubble expansion rate, $G$ is Newton's constant, $\rho$ is the energy density, $K=0,1,-1$ is the curvature constant of a maximally symmetric $3$-space, $a$ is the scale factor and $\Lambda$ is the cosmological constant. The asymptotic value of $\rho$ at late times can be set to zero by redefinition of $\Lambda$. In the standard cosmology, $\rho$ then includes energy densities of radiation ($\propto 1/a^4$) and pressure-less matter ($\propto 1/a^3$). The fact that all but $\Lambda$ decay as the universe expands is the source of the cosmological constant problem. The present paper does not intend to solve the cosmological constant problem and we simply assume that $\Lambda$ has the observed value. The slowest decaying component on the right-hand side of the Friedmann equation is the spatial curvature term $-3K/a^2$ and is the source of the flatness problem in the standard cosmology.

Inflation, once it occurs, makes $\rho$ almost constant for an extended period in the early universe so that even the curvature term decays faster than $\rho$. The initial condition of the standard cosmology is thus set at the end of inflation in such a way that the curvature term is sufficiently smaller than $8\pi G\rho$. Subsequently, the ratio of the curvature term to $8\pi G\rho$ grows but the initial value of the ratio at the end of inflation is so small that the universe reaches the current epoch before the ratio becomes order unity. This is how inflation solves the flatness problem.

If a theory of quantum gravity predicts that the ratio $(3K/a^2)/(8\pi G\rho)$ be sufficiently small at the beginning of the universe then this could be an alternative solution to the flatness problem. The purpose of the present paper is to propose such a solution based on the projectable version of Ho\v{r}ava-Lifshitz (HL) gravity~\cite{Horava:2009uw,Sotiriou:2009bx}, which has recently been proved to be renormalizable~\cite{Barvinsky:2015kil,Barvinsky:2017zlx} and thus is a good candidate for a quantum gravity theory. Since our proposal is solely based on a fundamental principle called the anisotropic scaling, which is respected by all versions of the HL theory, it is expected that the same idea can be implemented in other versions of HL gravity.

One of the fundamental principles of HL gravity is the so-called {\it anisotropic scaling}, or Lifshitz scaling,
\begin{equation}
 t \to b^z t\,, \quad \vec{x} \to b \vec{x}\,, 
  \label{eqn:anisotropic-scaling}
\end{equation}
where $t$ is the time coordinate, $\vec{x}$ are the spatial coordinates and $z$ is a number called dynamical critical exponent. In $3+1$ dimensions the anisotropic scaling with $z=3$ in the ultraviolet (UV) regime is the essential reason for renormalizability. The anisotropic scaling with $z=3$ also leads to a novel mechanism of generating scale-invariant cosmological perturbations, solving the horizon problem without inflation~\cite{Mukohyama:2009gg}.

In the context of quantum cosmology, the initial conditions of the universe are typically set by quantum tunneling described by an instanton, i.e.~a classical solution to some Euclidean equations of motion with suitable boundary conditions. In relativistic theories, where $z=1$, quantum tunneling is thought to be dominated by an $O(4)$ symmetric instanton, implying that $T=L$, where $T$ and $L$ are the Euclidean time and length scales, respectively. After analytic continuation to the real time evolution, this causes the flatness problem unless inflation follows.

Setting $z=3$, however, the story is completely different. An instanton should lead to $T\propto L^3$ and thus
\begin{equation}
 T\simeq M^2L^3\,, \label{eqn:T-LLL}
\end{equation}
where $T$ and $L$ are again the Euclidean time and length scales, respectively, and $M$ is the scale above which the anisotropic scaling (\ref{eqn:anisotropic-scaling}) with $z=3$ becomes important. If the theory is UV complete then the scaling (\ref{eqn:T-LLL}) is expected to apply to any kind of instantons deep in the UV regime, i.e. for $L\ll 1/M$. If the size of the instanton $L$ is indeed much smaller than $1/M$ then this implies that $T\ll L$ and thus the instanton has a highly anisotropic shape. We thus call this kind of instanton an {\it anisotropic instanton}. If the creation of the universe is dominated by a small anisotropic instanton then in the real time universe after analytic continuation, the spatial curvature length scale will be much greater than the cosmological time scale. In this way the anisotropic instanton may solve the flatness problem without inflation.

The rest of the present paper is organized as follows. In Section~\ref{HLgrav} we review projectable HL theory, obtaining the equivalent of Friedmann's equation (\ref{fried0}) in this theory. New curvature-dependent terms are found, which will be essential for the solution to the flatness problem proposed here. In Section~\ref{sec:noboundary} we examine a quantum state inspired by the no-boundary proposal: the idea that the universe nucleated from nothing, as represented by Euclidean evolution replacing the Big Bang singularity. We find that under anisotropic scaling and the semi-classical evolution of HL theory, the curvature is sufficiently suppressed to solve the flatness problem without the need for inflation. The solution may be more general than the concrete model presented here, as argued in Section~\ref{general}, where we show that on dimensional grounds we can always predict the modifications to  (\ref{fried0}) from the modified dispersion relations of the theory. Together with equipartition of energy at the initial point, evolution in this regime enforces the necessary suppression of the curvature. In Appendices \ref{app:perturbation} and \ref{sec:evolution} we discuss generation of scale-invariant perturbations and evolution after instanton based on the concrete setup of Section~\ref{sec:noboundary}. Appendix~\ref{app:moregeneral} then discusses further generalization of the already general scenarios of Section~\ref{general}.

\section{Projectable HL gravity}\label{HLgrav}

The basic variables of the projectable version of HL gravity are:
\begin{equation}
 \mbox{lapse}: N(t)\,,\ 
  \mbox{shift}: N^i(t,\vec{x})\,,\ 
   \mbox{3d metric}: g_{ij}(t,\vec{x})\, \label{eqn:lapse-shift-metric}\,.
\end{equation}
The theory respects the so-called foliation preserving diffeomorphisms,
\begin{equation}
 t \to t'(t)\,,\quad \vec{x}\to \vec{x}'(t,\vec{x})\,.
\end{equation}
Adopting the notation of \cite{Mukohyama:2010xz}, the action of the gravity sector is then given by 
\begin{equation}
 I_g = \frac{M_{\rm Pl}^2}{2}\int Ndt\sqrt{g}d^3\vec{x}\left(K^{ij}K_{ij}-\lambda K^2 -2\Lambda+R+L_{z>1}\right)\,,
\end{equation}
where
\begin{eqnarray}
 \frac{M_{\rm Pl}^2}{2}L_{z>1} & = &
  ( c_1D_iR_{jk}D^iR^{jk}+c_2D_iRD^iR+c_3R_i^jR_j^kR_k^i\nonumber\\
 & & +c_4RR_i^jR_j^i + c_5R^3)
  + (c_6R_i^jR_j^i+c_7R^2)\,.
\end{eqnarray}
Here, $K_{ij}=(\partial_t g_{ij}-D_iN_j-D_jN_i)/(2N)$ is the extrinsic curvature of the constant $t$ hypersurfaces, $K^{ij}=g^{ik}g^{jl}K_{kl}$, $K=g^{ij}K_{ij}$, $N_i=g_{ij}N^j$, $g^{ij}$ is the inverse of $g_{ij}$, $D_i$ and $R_i^j$ are the covariant derivative and the Ricci tensor constructed from $g_{ij}$, $R=R_i^i$ is the Ricci scalar of $g_{ij}$, $M_{\rm Pl}=1/\sqrt{8\pi G}$ is the Planck scale, and $\lambda$ and $c_{n}$ ($n=1,\cdots,7$) are constants.

In HL gravity, as already stated in (\ref{eqn:lapse-shift-metric}), a spacetime geometry is described by a family of spatial metrics parameterized by the time coordinate $t$, together with the lapse function and the shift vector. The 3D space at each $t$ can have non-trivial topology and may consist of several connected pieces, $\Sigma_{\alpha}$ ($\alpha=1,\cdots$), each of which is disconnected from the others. In this situation, we have a common lapse function and a set of shift vectors and a set of spatial metrics parameterized by not only (continuous) $t$ but also (discrete) $\alpha$, as
\begin{equation}
  N^i = N_{\alpha}^i(t,\vec{x})\,, \quad
  g_{ij} = g^{\alpha}_{ij}(t,\vec{x})\,, \quad (\vec{x}\in \Sigma_{\alpha})\,.
\end{equation}
The equation of motion for $N(t)$ then leads to a global Hamiltonian constraint of the form, 
\begin{equation}
 \sum_{\alpha} \int_{\Sigma_{\alpha}}d^3\vec{x}\, \mathcal{H}_{g\perp}=0\,,
  \label{eqn:Hamiltonian-constraint}
\end{equation}
where
\begin{equation}
 \mathcal{H}_{g\perp} = \frac{M_{\rm Pl}^2}{2}\sqrt{g}(K^{ij}K_{ij}-\lambda K^2+2\Lambda-R-L_{z>1})\,.
\end{equation}
Because of the summation over mutually disconnected pieces of the space $\{\Sigma_{\alpha}\}$ in (\ref{eqn:Hamiltonian-constraint}), 
\begin{equation}
 \int_{\Sigma_{\alpha}}d^3\vec{x}\, \mathcal{H}_{g\perp} \ne 0
\end{equation}
is possible, provided that the sum of them over all $\alpha$ is zero. Therefore, if we are interested in a universe in one of $\{\Sigma_{\alpha}\}$ then there is neither a local nor a global Hamiltonian constraint that needs to be taken into account. On the other hand, the equation of motion for $N^i(t,\vec{x})$ and $g_{ij}(t,\vec{x})$ are local and thus must be imposed everywhere. The absence of a Hamiltonian constraint introduces an extra component that behaves like dark matter~\cite{Mukohyama:2009mz,Mukohyama:2009tp}, as we shall see below explicitly for a homogeneous and isotropic universe.

We now consider a homogeneous and isotropic universe in each connected piece of the space $\Sigma_{\alpha}$ ($\alpha=1,\cdots$), described by 
\begin{equation}
 N_{\alpha}^i = 0\,, \quad g^{\alpha}_{ij} = a_{\alpha}(t)^2\Omega_{ij}\,,
\end{equation}
where $\Omega^{\alpha}_{ij}$ is the metric of the maximally symmetric three-dimensional space with the curvature constant $K_{\alpha}=0,1,-1$ and the Riemann curvature $R^{ij}_{\ \ kl}[\Omega^{\alpha}]=K_{\alpha}(\delta^i_k\delta^j_l-\delta^i_l\delta^j_k)$. The action is then
\begin{eqnarray}
 I_g & = & 3 M_{\rm Pl}^2 \int Ndt \sum_{\alpha}\int_{\Sigma_{\alpha}}d^3\vec{x}
    a_{\alpha}^3 \mathcal{L}_{\alpha}\,, \\
 \mathcal{L}_{\alpha} & = & 
   \frac{1-3\lambda}{2}H_{\alpha}^2 + \frac{\alpha_3 K_{\alpha}^3}{3a_{\alpha}^6}+\frac{\alpha_2K_{\alpha}^2}{a_{\alpha}^4}+\frac{K_{\alpha}}{a_{\alpha}^2}-\frac{\Lambda}{3}\,,\nonumber
\end{eqnarray}
where $H_{\alpha}=(\partial_t a_{\alpha})/(Na_{\alpha})$, $\alpha_2=4(c_6+3c_7)/M_{\rm Pl}^2$ and $\alpha_3=24(c_3+3c_4+9c_5)/M_{\rm Pl}^2$. The variation of the action with respect to $a_{\alpha}$ leads to the dynamical equation,
\begin{equation}
 \frac{3\lambda-1}{2}\left(2\frac{\partial_t H_{\alpha}}{N}+3H_{\alpha}^2\right)
  = \frac{\alpha_3K_{\alpha}^3}{a_{\alpha}^6}
  + \frac{\alpha_2K_{\alpha}^2}{a_{\alpha}^4}
  - \frac{K_{\alpha}}{a_{\alpha}^2} + \Lambda\,.
\end{equation}
Integrating this equation once, we obtain
\begin{equation}
 \frac{3(3\lambda-1)}{2}H_{\alpha}^2 = \frac{C_{\alpha}}{a_{\alpha}^3}
  - \frac{\alpha_3K_{\alpha}^3}{a_{\alpha}^6}
  - \frac{3\alpha_2K_{\alpha}^2}{a_{\alpha}^4}
  - \frac{3K_{\alpha}}{a_{\alpha}^2} + \Lambda\,,
\end{equation}
where $C_{\alpha}$ is an integration constant. The first term on the right-hand side behaves like a pressureless dust and thus is called {\it dark matter as integration constant}~\cite{Mukohyama:2009mz,Mukohyama:2009tp}. The equation of motion for $N(t)$ then leads to the global Hamiltonian constraint of the form (\ref{eqn:Hamiltonian-constraint}). For example, if $K_{\alpha}=1$ for ${}^{\forall}\alpha$ then the global Hamiltonian constraint is simply
\begin{equation}
 \sum_{\alpha} C_{\alpha} = 0\,.
\end{equation}
For the reason already explained in the previous paragraph, we do not need to consider this equation, if we are interested in a universe in one of $\{\Sigma_{\alpha}\}$.

\section{Anisotropic instanton}
\label{sec:noboundary}

As we have shown in the previous section, a homogeneous and isotropic universe in the projectable HL gravity is described by
\begin{equation}
 \frac{3(3\lambda-1)}{2}H^2 = \frac{C}{a^3}
  - \frac{\alpha_3K^3}{a^6}
  - \frac{3\alpha_2K^2}{a^4}
  - \frac{3K}{a^2} + \Lambda\,. \label{eqn:FriedmannEQ}
\end{equation}
Here, the subscript $\alpha$ has been suppressed. For simplicity, we set $\alpha_2=0$ and $\Lambda=0$ giving
\begin{equation}
 \frac{3(3\lambda-1)}{2}H^2 = \frac{C}{a^3} - \frac{\alpha_3K^3}{a^6} - \frac{3K}{a^2}\,.
\end{equation}
We assume that there is a UV fixed point of the renormalization group (RG) flow with a finite value of $\lambda$ larger than $1$, as in the case of $2+1$ dimensions~\cite{Barvinsky:2017kob}. Since we are interested in quantum tunneling in the UV, it is ideal to set $\lambda$ to a constant value ($>1$) at the UV fixed point. However, since the RG flow in $3+1$-dimensions has not yet been investigated, we shall consider $\lambda$ as a free parameter ($>1$). We shall adopt units in which $M_{\rm Pl}=1$. 

\begin{figure*}
	\centering
	\begin{subfigure}[t]{0.5\linewidth}
		\centering
		\includegraphics[height=2.0in]{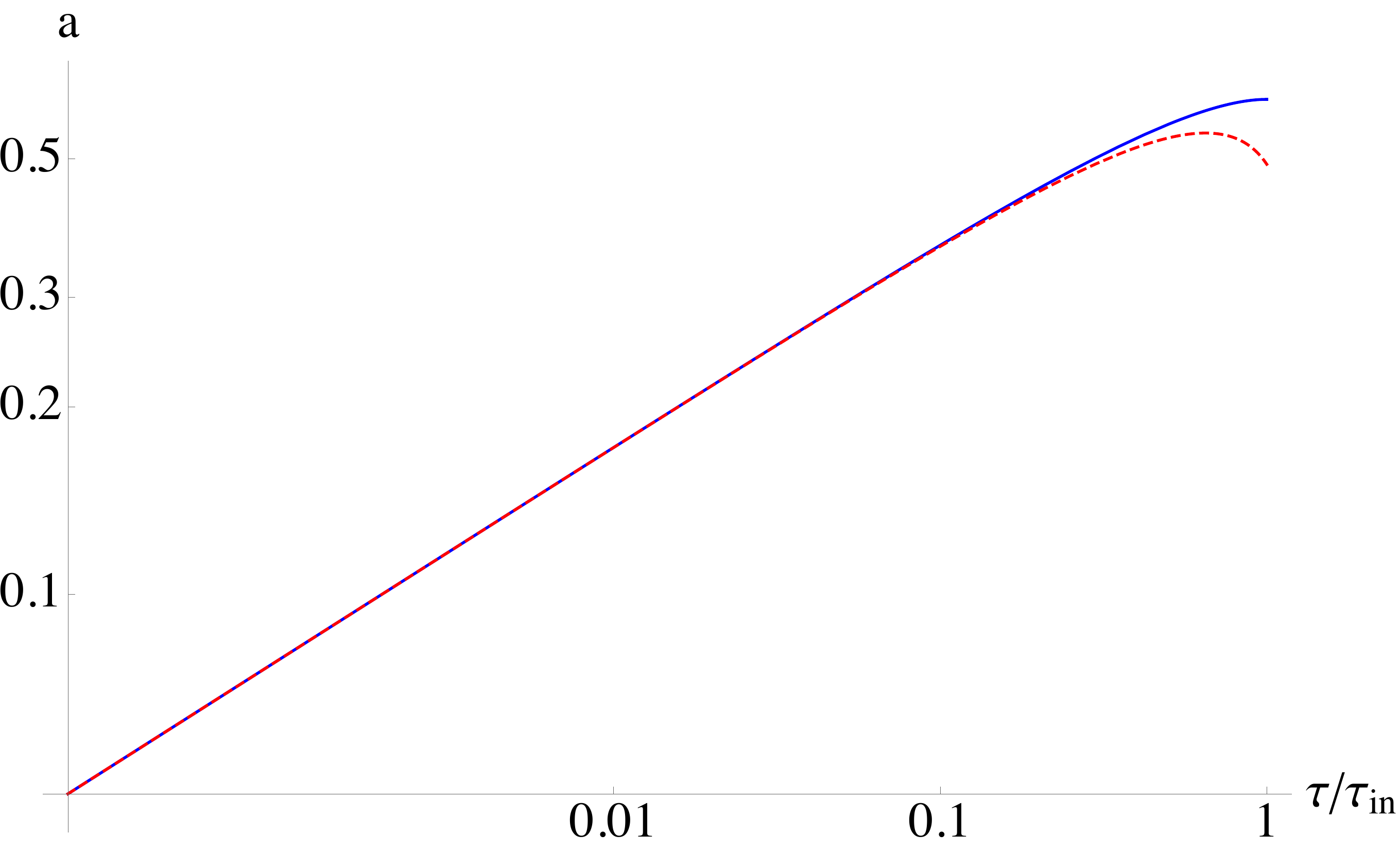}
	\end{subfigure}%
	~
	\begin{subfigure}[t]{0.5\linewidth}
		\centering
		\includegraphics[height=2.0in]{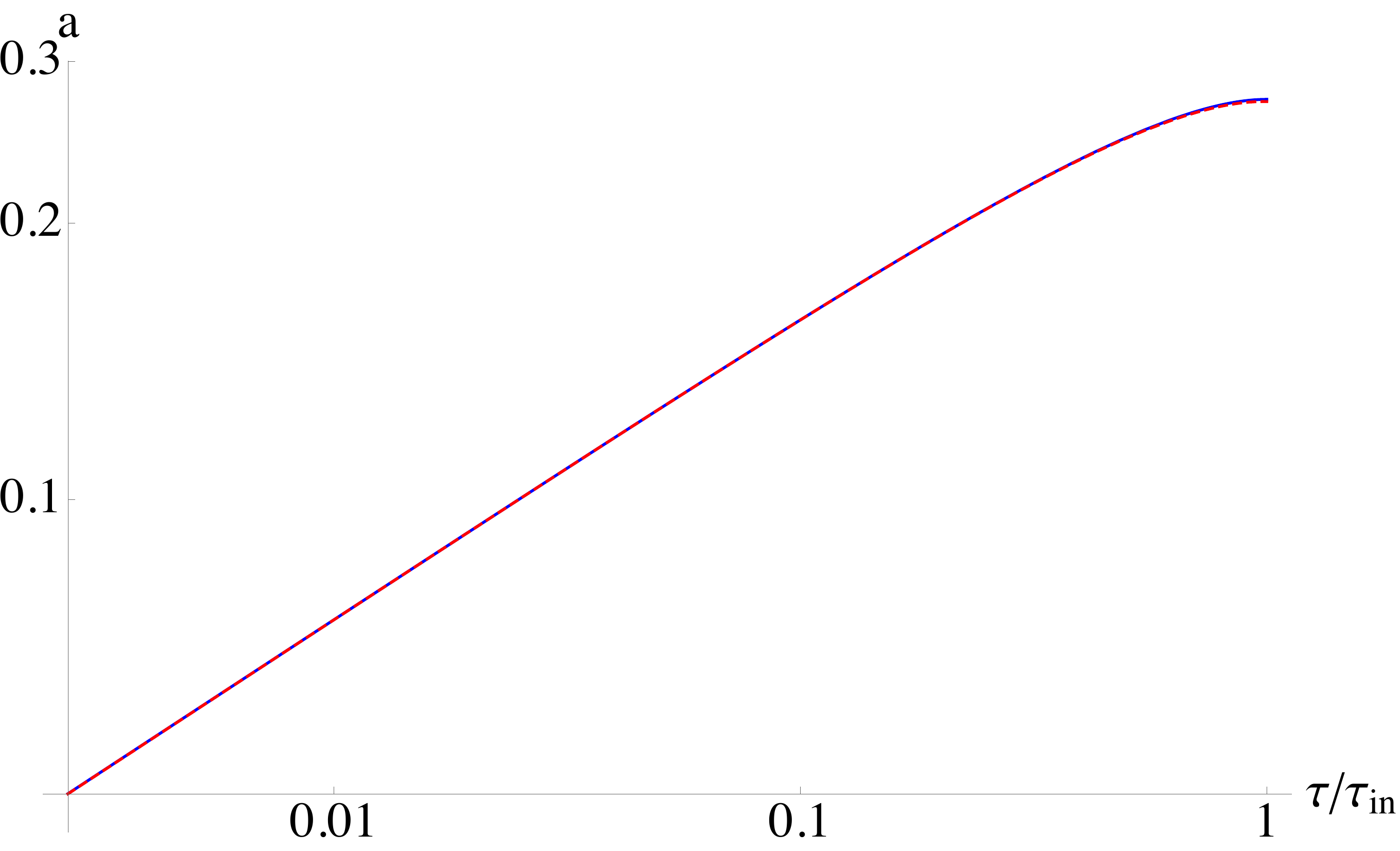}
	\end{subfigure}
	\caption{Loglog plots of $a$ vs. $\tau/ \tau_{\rm in}$ in solid blue with the analytic solution (\ref{eqn:a-approx-nearnb}) superimposed in dashed red. We have $\lambda = 2$, $\alpha_3 = 1$, $\alpha_2 = 0$ for both plots, however on the left we have $C = 6$ while on the right $C = 50$. This confirms the validity of the analytic solution in the large C limit. The figures were obtained by solving equation (\ref{eqn:FriedmannEQ}) numerically from $\tau = 10^{-4}$ using the small $\tau$ expansion until  $\partial_\tau a|_{\tau=\tau_{\rm in}} \approx 10^{-22}$.}
	\label{fig:numericalsol}
\end{figure*}

Hereafter in this section, we consider the creation of a closed ($K=1$) universe. Switching to Euclidean time $\tau=i\int^t N(t')dt'+const.$, we obtain
\begin{equation}
 \frac{3(3\lambda-1)}{2}\frac{(\partial_\tau a)^2}{a^2} =
  - \frac{C}{a^3} + \frac{\alpha_3}{a^6} + \frac{3}{a^2}\,. \label{eqn:EuclideanFriedmannEQ}
\end{equation}
Supposing that $a\to +0$ as $\tau\to +0$, the leading behavior of $a$ for small $\tau$ is $a\simeq a_1\tau^{1/3}$, where $a_1$ is a constant. Hence, expanding $a$ around $\tau=0$ as
\begin{equation}
 a = a_1\tau^{1/3} + a_2\tau^{2/3} + a_3\tau + \cdots\,,
\end{equation}
and plugging this into the Euclidean equation of motion (\ref{eqn:EuclideanFriedmannEQ}), we obtain
\begin{equation}
 a_1 = \left(\frac{6\alpha_3}{3\lambda-1}\right)^{1/6}, \ 
  a_2 = 0\,,\ 
  a_3 = \frac{3\alpha_2}{10}\sqrt{\frac{6}{\alpha_3(3\lambda-1)}}\,.
\end{equation}
By using this formula, it is easy to solve (\ref{eqn:EuclideanFriedmannEQ}) numerically from $\tau=\epsilon$ towards larger $\tau$, where $\epsilon$ is a small positive number. The solution is unique for a given value of the integration constant $C$ as we have already fixed another integration constant corresponding to a constant shift of $\tau$. Some numerical solutions are shown in figure \ref{fig:numericalsol}. For a positive $\alpha_3$ and a large enough positive $C$, one finds that $\partial_{\tau} a$ vanishes at a finite value of $\tau$, which we call $\tau_{\rm in}$, i.e.
\begin{equation}
 \partial_\tau a|_{\tau=\tau_{\rm in}} = 0\,. \label{eqn:def-tauc}
\end{equation}
The Lorentzian evolution of the universe after the quantum tunneling is then obtained by Wick rotating the Euclidean solution at $\tau=\tau_{\rm in}$ as $\tau=\tau_{\rm in}+i\int^t N(t')dt'$, meaning that the instanton is represented by the solution in the range $\epsilon\leq\tau\leq\tau_{\rm in}$ with $\epsilon\to +0$. 
The contribution of the connected piece of the space of interest to the Euclidean action $iI_g$ is then
\begin{eqnarray}
 S_{\rm E} & = & 6\pi^2 \lim_{\epsilon\to +0}\int_{\epsilon}^{\tau_{\rm in}} d\tau \left[\frac{1-3\lambda}{2}a(\partial_{\tau}a)^2 - \frac{\alpha_3}{3a^3}-a\right]\nonumber\\
 & = & 6\pi^2 \lim_{\epsilon\to +0}\int_{\epsilon}^{\tau_{\rm in}} d\tau \left[ \frac{C}{3}-\frac{2\alpha_3}{3a^3}-2a\right]\,, \label{eqn:Euclideanaction}
\end{eqnarray}
where we have used the equation of motion (\ref{eqn:EuclideanFriedmannEQ}).

\begin{figure}
  \begin{center}
   \includegraphics[height=2.0in]{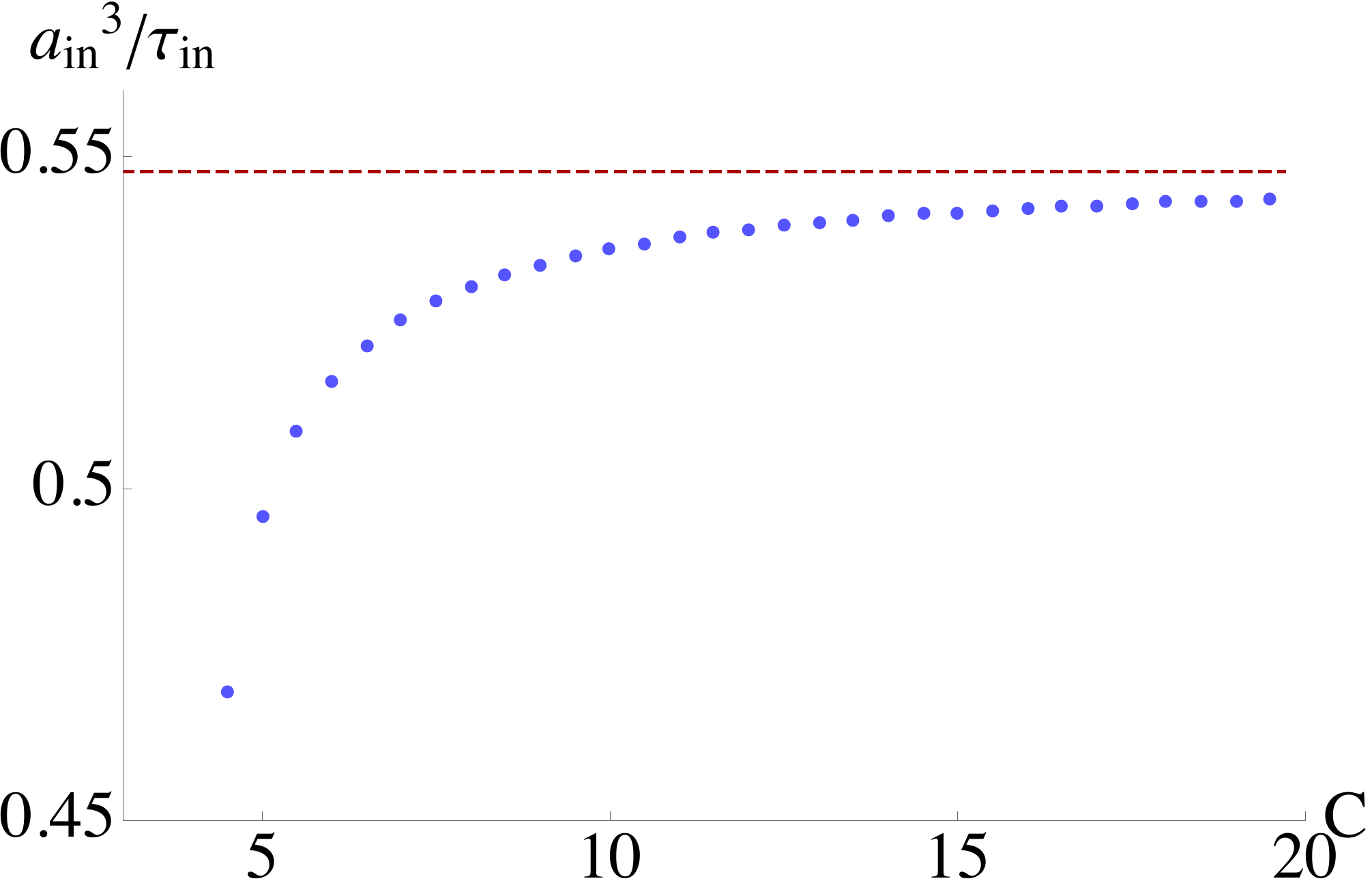}
   \caption{The plot shows $a_{\rm in}^3/\tau_{\rm in}$ as a function of $C$ and confirms the expected analytic scaling behavior in the large C limit shown in dashed red. To obtain the plot, we kept $\lambda = 2$, $\alpha_3 = 1$, $\alpha_2 = 0$ and integrated the Euclidean equation of motion from $\tau = 0$ to $\tau = \tau_{\rm in}$ for various values of the integration constant $C$.}
   \label{fig:varyC}
  \end{center}
 \end{figure}

For large positive $C$, we expect $a$ to be small in the whole interval $0\leq\tau\leq\tau_{\rm in}$. Hence in this limit we can safely ignore the last term on the right-hand side of (\ref{eqn:EuclideanFriedmannEQ}):
\begin{equation}
 \frac{3(3\lambda-1)}{2}\frac{(\partial_\tau a)^2}{a^2} \simeq 
  - \frac{C}{a^3} + \frac{\alpha_3}{a^6}\,.
\end{equation}
We then have an approximate analytic solution given by
\begin{equation}
 \sqrt{\frac{2}{3(3\lambda-1)}}\tau \simeq \frac{2\sqrt{\alpha_3}}{3C}\left(1 - \sqrt{1-\frac{C}{\alpha_3}a^3}\right)\,,
\end{equation}
or equivalently 
\begin{equation}
 a \simeq \left[ 3\sqrt{\alpha_3}\mathcal{T} - \frac{9}{4}C\mathcal{T}^2\right]^{1/3}\,, \quad
  \mathcal{T} = \sqrt{\frac{2}{3(3\lambda-1)}}\tau\,. \label{eqn:a-approx-nearnb}
\end{equation}
As a result, we have
\begin{equation}
 \tau_{\rm in} \simeq \frac{2\sqrt{\alpha_3}}{3C}\sqrt{\frac{3(3\lambda-1)}{2}}\,,\quad
  a_{\rm in} \simeq \left(\frac{\alpha_3}{C}\right)^{1/3}\,, \label{eqn:tauc-ac-approx}
\end{equation}
where $a_{\rm in}\equiv a(\tau_{\rm in})$. This implies that
\begin{equation}
 \frac{a_{\rm in}^3}{\tau_{\rm in}} \simeq \sqrt{\frac{3\alpha_3}{2(3\lambda-1)}} = \mbox{const}\,. \label{eqn:ac3_over_tauc}
\end{equation}
Since we have set $K=1$, the scale factor $a$ has the dimension of length. For a positive $\alpha_3$ and a large positive value of $C$, $a_{\rm in}\equiv a(\tau_{\rm in})$ is small as seen in (\ref{eqn:tauc-ac-approx}). As expected from the scaling argument (\ref{eqn:T-LLL}) in the Introduction and as confirmed numerically in figure \ref{fig:varyC}, we have the scaling relation (\ref{eqn:ac3_over_tauc}). These results support the claim that a small anisotropic instanton may solve the flatness problem in HL gravity.

 To see if the small instanton dominates the creation of the universe, we need to estimate the tunneling rate, which in the regime of validity of the semi-classical approximation, is given by the exponential of the Euclidean action (\ref{eqn:Euclideanaction}). This however turns out to be a difficult task. First, both the (Euclidean) extrinsic curvature $K^i_{\rm E\, j}=\delta^i_j\partial_{\tau}\ln a$ and the spatial curvature $R^i_{\ j}=2\delta^i_j/a^2$ diverges in the limit $\tau\to +0$, indicating that the semi-classical description should break down near $\tau=0$. We are thus unable to rely on the semi-classical formula for the tunneling rate. Indeed, the dominant term in the integrand of (\ref{eqn:Euclideanaction}) for small $\tau$ is $\propto \alpha_3/a^3\propto\sqrt{\alpha_3(3\lambda-1)}/\tau$, whose integral over the small $\tau$ region exhibits a divergence of order $\sqrt{\alpha_3(3\lambda-1)}\ln\epsilon$. Thus the quantum state employed in this paper, while inspired by the no-boundary proposal, does not have a regular beginning. Quantum effects such as the RG flow of coupling constants might somehow ameliorate the logarithmic divergence but this is beyond the scope of the present paper. Second, based on a formulation of the Lorentzian path integral for quantum cosmology, it was recently suggested that the semi-classical formula for the tunneling rate may have to be drastically modified~\cite{Feldbrugge:2017kzv,Feldbrugge:2017fcc,Feldbrugge:2017mbc}. This may pose some doubts on the no-boundary proposal~\cite{Hawking:1981gb,Hawking:1983hj,Hartle:1983ai,Hartle:2008ng,DiazDorronsoro:2017hti} in general relativity. It is certainly worthwhile investigating whether a similar argument applies to HL gravity or not.

The Euclidean solution that we found is unique up to an integration constant $C$ and a physically irrelevant, arbitrary shift of the origin of the Euclidean time coordinate $\tau$, as far as the homogeneous and isotropic ansatz with a positive three-dimensional curvature is adopted. Therefore one can easily show that the scaling (\ref{eqn:T-LLL}) holds for large $C$, independently from the boundary condition near $a=0$. This is because the (Euclidean) time scale $T$ and the length scale $L$ at $\tau=\tau_{\rm in}$ can be defined locally, without referring to the behavior of the solution away from $\tau=\tau_{\rm in}$, as
\begin{equation}
 T \sim \left|\dot{H}(\tau_{\rm in})\right|^{-1/2}\,,\quad
  L \sim \left|\frac{K}{a_{\rm in}^2}\right|^{-1/2}\,. \label{eqn:T-L-localdef}
\end{equation}
Since the equation of motion implies that
\begin{equation}
 \left|\dot{H}(\tau_{\rm in})\right|
  \sim \frac{1}{M^4}\left|\frac{K}{a_{\rm in}^2}\right|^3\,,
\end{equation}
it follows that
\begin{equation}
 T \sim M^2 L^3\,, \label{eqn:anisotropic-instanton-scaling}
\end{equation}
where $M\equiv \alpha_3^{-1/4}$ is the momentum scale above which the anisotropic scaling with $z=3$ and thus the curvature cubic term in (\ref{eqn:FriedmannEQ}) become important. This is exactly what we have expected in (\ref{eqn:T-LLL}) from general arguments. Because of the uniqueness of the Euclidean solution, this scaling holds for $L\ll 1/M$, independently from the boundary condition and any physical conditions near $a=0$.

\section{General argument}\label{general}

Although we have proposed a concrete framework for solving the flatness problem within HL gravity, the arguments presented are more general and may be valid on purely dimensional grounds for any UV complete theory with an anisotropic scaling of spacetime. This can be suspected from the simple argument presented in Section~\ref{intro}, but we now take the dimensional argument further. All that we shall need from the concrete model presented are its dispersion relations (as in HL theory) and equipartition at the starting point (as imposed by the anisotropic instanton).

Let a general UV complete theory have modified dispersion relations for its massless particles (including gravitons) of the form:
\begin{equation}\label{MDR}
 E^2= M^2f(p^2/M^2)\,, 
\end{equation}
where $f$ is a smooth function with the following asymptotic behavior,
\begin{equation}\label{MDR-ap}
 f(x) = \left\{\begin{array}{ll}
	 x\,, & (0\leq x\ll 1)\\
		x^z\,, & (x\gg 1)
	       \end{array}\right.\,,
\end{equation}
and the momentum scale $M$ may be taken to be of the order of the Planck scale or not. 
This is a Hamiltonian constraint for particles, so we may expect that in a FLRW setting a corresponding Hamiltonian constraint for vacuum solutions will result from replacing $E^2\rightarrow H^2$ and $p^2\rightarrow |K|/a^2$. Even when such a constraint does not strictly exist (as is the case with the HL model), an effective one may be present, resulting in a Friedmann-like equation. On dimensional grounds we expect the corresponding Friedmann equation in vacuum to read:
\begin{equation}
H^2= \pm M^2f(|K|/a^2M^2)\, .
\end{equation}
The sign on the right-hand side may be either positive or negative and the following argument does not rely on the choice of the sign. Addition of matter energy density $\rho$ (or some component that stems from gravity but that behaves like matter, such as the term $C/a^3$ in (\ref{eqn:FriedmannEQ})) then leads to:
\begin{equation}\label{fried}
H^2=\frac{1}{3}\frac{\rho}{M_{\rm Pl}^2} \pm M^2f(|K|/a^2M^2)\,,
 \end{equation}
where we have assumed that the ratio between the effective gravitational constant for the homogeneous and isotropic cosmology and $(8\pi M_{\rm Pl}^2)^{-1}$ is (approximately) constant~\footnote{In the concrete example of the previous section, this ratio depends on $\lambda$ and thus is in general subject to running under the RG flow. However, as we have assumed the existence of a UV fixed point with finite $\lambda$, this assumption is justified.} and we have absorbed such a ratio into the definition of $\rho$. To complete the system we have to specify the second Friedmann equation (which indeed was the starting point for our concrete model), or alternatively, the conservation equation for $\rho$. Let us first assume conservation (this is in fact not needed and violations of energy conservation only refine and reinforce the argument: see Appendix \ref{app:moregeneral} for details). With a general equation of state $w=p/\rho$ we then have:
\begin{equation}\label{rho-eq}
\dot \rho + 3H(1+w)\rho=0\,,
\end{equation}
integrating into:
\begin{equation}\label{rho-sol}
\rho\propto \frac{1}{a^{3(1+w)}}\,.
\end{equation}
In our concrete model we have $z=3$ and $w=0$, but this set up is more general.

Let us now assume that at some time, deep in the UV regime far beyond the scale $M$, the Lorentzian signature universe is created, after which it is subject to (semi-) classical evolution. We assume that the theory we are considering is UV complete, so there is no need to fear going beyond the scale $M$. This ``initial time'' of creation can be seen as the result of tunneling from vacuum, via an instanton, similar to our concrete model, or it can be the result of any other process, e.g.~a phase transition from a disordered quantum geometry. The point is that the Universe undergoes a transition {\it into} (semi-) classical evolution in the UV complete theory at a density $\rho_{\rm in}$, assumed to be $\rho_{\rm in}\gg M_{\rm Pl}^2M^2$.

Let us now also assume that an equipartition principle is in action, that is, we assume roughly equal amounts of energy for different types of contributions that enter the Hamiltonian. In our setting there are just two contributions: matter (with a general equation of state $w$) and curvature. Curvature can be seen as a fluid with energy density:
\begin{equation}\label{rhoK}
\rho_K=\pm 3M_{\rm Pl}^2M^2f(|K|/a^2M^2)\,,
\end{equation}
and we can tweak this formula as appropriate, to contain the concrete model. Equipartition, then, implies:
\begin{equation}
\rho\approx \rho_K\sim\rho_{\rm in}\,,
\end{equation}
which is equivalent to the suppression of curvature $K/a^2$ derived from the anisotropic instanton presented in Section III. However, defined in terms of $\rho_K$ there is no suppression. Indeed $\rho\sim\rho_K\sim \rho_{\rm in}$ initially and the subsequent evolution takes care of the suppression. Whether we phrase things in terms of $K/a^2$ or $\rho_K$ the final result is the same. 

Let the curvature be measured by
\begin{equation}
\Omega_K=\frac{\rho_K}{\rho+\rho_K}\,.
\end{equation}
Using (\ref{rho-sol}) and (\ref{rhoK}) with \eqref{MDR-ap} we see that for $M^2 \ll \vert K \vert / a^2 \ll M^2 ( \rho_{\rm in} / M_{\rm Pl}^2M^2)^{1/z}$ or equivalently for $\rho_{z\to 1} \ll \rho \ll \rho_{\rm in}$, where
\begin{equation}
 \rho_{z\to 1} \sim \rho_{\rm in} \left(\frac{M_{\rm Pl}^2M^2}{\rho_{\rm in}}\right)^{\frac{3(1+w)}{2z}}\,, \label{eqn:def-rhozto1}
\end{equation}
we have:
\begin{equation}\label{flat-sol}
\Omega_K\propto a^{3(1+w)-2z}\,,
\end{equation}
whereas for $\vert K \vert / a^2 \ll M^2$ or equivalently for $\rho \ll \rho_{z\to 1}$ we have the standard flatness problem instability:
\begin{equation}\label{flat-prob}
\Omega_K\propto a^{3(1+w)-2}\,.
\end{equation}
So that $\Omega_K$ may be suppressed in the first stage of evolution we see that a necessary condition for solving the flatness problem in an expanding universe is:
\begin{equation}
z>\frac{3(1+w)}{2}\,.
\end{equation}
In our concrete model this is satisfied since $z=3$ and $w=0$, but in fact for the $z=3$ HL theory this would work with any $w<1$. With standard gravity (i.e. $z=1$) we would need $w<-1/3$, i.e. inflation. 

The above is a necessary but not sufficient condition. The exact condition will involve $M$ and $\rho_{\rm in}$ as well as $z$ and $w$. Assuming for simplicity the universe exits the UV phase around $\vert K \vert / a^2 \sim M^2$ to enter a standard hot big bang model, then curvature must be suppressed at this time by:
\begin{equation}
\Omega_K \ll \Omega_{\rm sup} \equiv z_{\rm eq}\left(\frac{T_{\rm CMB}}{M_{\rm Pl}} \right)^2 
\left(\frac{M_{\rm Pl}^4}{\rho_{z\to 1}} \right)^{\frac{1}{2}}\,,
\end{equation}
where $T_{\rm CMB}$ is the present temperature of the cosmic microwave background, we have used (\ref{flat-prob}) with $w=1/3$ and $0$ before and after matter radiation equality, and $z_{\rm eq}$ is the redshift of matter radiation equality. If $\rho_{z\to 1} \sim M_{\rm Pl}^4$, with standard assumptions we have roughly $\Omega_{\rm sup}\sim 10^{-60}$, as is well known. 

In order to obtain this suppression while $\rho_{z\to 1} < \rho< \rho_{\rm in}$ we should therefore impose the condition:
\begin{equation}\label{condition}
\frac{\rho_{\rm in}}{\rho_{z\to 1}}\gg \Omega_{\rm sup}^{-\frac{3(1+w)}{2z-3(1+w)}}\,,
\end{equation}
where we have used (\ref{flat-sol}) in conjunction with $\rho$ conservation (and solution (\ref{rho-sol})), even though the latter is not strictly necessary. Expressing $\rho_{z\to 1}$ and $\Omega_{\rm sup}$ in terms of $\rho_{\rm in}$ and $M$, this translates to
\begin{equation}
\frac{\rho_{\rm in}}{M_{\rm Pl}^2M^2} \gg \left[ \frac{1}{z_{\rm eq}} \frac{M_{\rm Pl}M}{T_{\rm CMB}^2} \right]^{\frac{4z}{2z-3(1+w)}} \; .
\label{rhoinbyM}
\end{equation}
Eq.~\eqref{rhoinbyM} is the general condition for solving the flatness problem in the vast class of models considered here. 
For the concrete model proposed in this paper ($z=3$, $w=0$), if we set $M = M_{\rm Pl}$ for concreteness then we have:
\begin{equation}
\frac{\rho_{\rm in}^{1/4}}{M_{\rm Pl}} \gg \frac{1}{z_{\rm eq}} \left(\frac{M_{\rm Pl}}{T_{\rm CMB}} \right)^2
\sim 10^{58}
\; .
\label{rhoinMpl-general}
\end{equation}

Eq.~(\ref{rhoinbyM}) establishes the general condition for a solution of the flatness problem in general UV complete theories with anisotropic scaling. In summary, they must start operating sufficiently above the Planck scale and satisfy equipartition in some form at this initial point. This applies to our concrete model with a starting point defined by an anisotropic instanton. However, the formal mechanism is more general.

\section{Summary and discussions}

We have proposed a possible solution to the cosmological flatness problem without relying on inflation. To do so we have made use of the renormalizable theory of gravity called Ho\v{r}ava-Lifshitz (HL) gravity. We further assumed that the initial condition of the universe respects the so-called anisotropic scaling (\ref{eqn:anisotropic-scaling}), with $z=3$ which is the minimal value that guarantees renormalizability of HL gravity. Because of this scaling, any physical system in the deep ultraviolet (UV) regime tends to possess the scaling property $T\simeq M^2L^3$, where $T$ and $L$ are the time scale and the length scale of the system and $M$ is the momentum scale characterizing the anisotropic scaling. If the universe started in the deep UV regime then the initial condition is expected to satisfy this scaling property with $L\ll 1/M$, meaning that the curvature length scale of the universe is much longer than the expansion time scale. This is exactly what is needed for solving the flatness problem.

Based on the projectable version of the HL theory for concreteness, we have found a family of instanton solutions parameterized by an integration constant $C$. This family of solutions is unique under the FLRW ansatz for the pure gravity system, i.e. without any matter fields, for a given set of parameters in the action. For positive and large enough $C$, the spatial size $a_{\rm in}$ and the (Euclidean) temporal size $\tau_{\rm in}$ of the instanton are decreasing functions of $C$. We confirmed the scaling relation $a_{\rm in}^3/\tau_{\rm in}\simeq const.$ in the large $C$ limit, both numerically and analytically. Moreover, by defining $T$ and $L$ locally at $\tau=\tau_{\rm in}$ through $\dot{H}$ and $K/a^2$ as in (\ref{eqn:T-L-localdef}), we have seen that the scaling $T\sim M^2 L^3$ holds independently from the boundary condition and any physical conditions near $a=0$. We call those instantons with anisotropy in $4$-dimensional Euclidean spacetime (but with isotropy in $3$-dimensional space) {\it anisotropic instantons}. The anisotropic instanton provides a concrete example of a physical system that realizes the scaling property $T\simeq M^2L^3$ and thus may solve the flatness problem in cosmology.

We have also given a more general argument for the solution of the flatness problem presented here, based on the assumption of equipartition among different contributions of energy density to the Hamiltonian of the system. The equipartition between the highest time derivative term and the highest spatial gradient term can be considered as a restatement of the anisotropic scaling and thus is expected to be universally applicable to many physical systems in any possible UV complete theories with anisotropic scaling. We showed that a large class of theories and cosmological models satisfying this property will be free from the flatness problem. The flatness of the Universe is then an expression of the fact that the Universe started deep in the UV regime, and of this scaling property of quantum gravity.

\begin{acknowledgements}
The work of S. F. B. is supported in part by a grant from the Studienstiftung des Deutschen Volkes. S. F. B. and A. C. are grateful for the hospitality of the Yukawa Institute of Theoretical Physics where some of this work was completed.
The work of A.C. was undertaken in part as an overseas researcher under a Short-Term Fellowship of the Japan Society for the Promotion of Sciences.
The work of J. M. was supported by  an STFC consolidated  grant and the John Templeton Foundation.
The work of S. M. was supported by Japan Society for the Promotion of Science (JSPS) Grants-in-Aid for Scientific Research (KAKENHI) No. 17H02890, No. 17H06359, No. 17H06357. 
The work of S. M. and Y. W. were supported by World Premier International Research Center Initiative (WPI), MEXT, Japan.
The work of R. N. was supported by the Natural Sciences and Engineering Research Council of Canada and by the Lorne Trottier Chair in Astrophysics and Cosmology at McGill. 
The work of Y. W. was supported by JSPS Grant-in-Aid for Scientific Research No.\,16J06266 and by the Program for Leading Graduate Schools, MEXT, Japan. 
\end{acknowledgements}

\appendix

\section{Scale-invariant perturbation}
\label{app:perturbation}

In the projectable Ho\v{r}ava-Lifshitz gravity, there are three physical degrees of freedom: two from the tensor graviton and one from the scalar graviton. Actually, one can consider the scalar graviton as a perturbation of the ``dark matter as integration constant'', i.e. the $C/a^3$ term in (\ref{eqn:FriedmannEQ}). In other words, the ``dark matter as integration constant'' is a coherent condensate of scalar gravitons. Both tensor and scalar gravitons obey the $z=3$ anisotropic scaling and thus it is expected that the quantum tunneling comes with scale-invariant cosmological perturbations of both of them, following exactly the same logic as the one proposed in \cite{Mukohyama:2009gg}.

After quantum tunneling, the universe is still in the UV regime and thus the stress-energy tensor of matter fields $T_{\mu\nu}$ does not have to satisfy the usual four-dimensional conservation equation, $\nabla^{\mu}T_{\mu\nu}\ne 0$, where $\nabla^{\mu}$ is the four-dimensional covariant derivative. In this situation, matter fields and the scalar graviton exchange energies~\cite{Mukohyama:2009mz}. It is therefore possible that the scale-invariant perturbations of the scalar graviton may be transferred to matter fields. As a result of such transfer processes a part of the coherent condensate of scalar gravitons, i.e. the ``dark matter as integration constant'', may be converted to a gas/dust of scalar graviton particles, which may also behave as dark matter. If energy densities in the matter sector are initially small compared with that in the ``dark matter as integration constant'' then the resulting perturbations after such transfer of energies will inevitably be almost scale-invariant and adiabatic.

\section{Evolution after instanton}
\label{sec:evolution}

In Sec.~\ref{sec:noboundary}, we have shown that for a large positive value of the \emph{dark matter as integration constant}, $C$, there is an instanton solution with the scaling properties (\ref{eqn:ac3_over_tauc}) and (\ref{eqn:anisotropic-instanton-scaling}). After the Lorentzian universe emerges as the analytic continuation of the instanton, this ``dark matter'' dominates the energy density of the universe for some time. For the subsequent evolution, we assume a minimal scenario as a demonstration in this appendix. Since $C$ interacts with matter (and thus in fact is not constant in its presence) \cite{Mukohyama:2010xz}, it can decay and populate the universe with matter and radiation some time after the instanton tunneling. Under the assumption that the continuity equation is respected in the matter sector, the relevant part of \eqref{eqn:FriedmannEQ} after the transition reads, with addition of matter and radiation,
\begin{equation}
\frac{3(3\lambda-1)}{2}H^2 = \frac{C}{a^3}
- \frac{3K}{a^2} 
+ \frac{1}{M_{\rm Pl}^2}\left(\rho_{\rm mat} + \rho_{\rm rad}\right)
+ \Lambda \; ,
\label{Friedmann-late}
\end{equation}
where $\rho_{\rm mat}$ and $\rho_{\rm rad}$ are the energy densities of radiation and matter, respectively, and we have absorbed the ratio between the effective gravitational constant for the homogeneous and isotropic cosmology and $(8\pi M_{\rm Pl}^2)^{-1}$ into the definition of $\rho_{\rm matter}$ and $\rho_{\rm rad}$. We have recovered the cosmological constant $\Lambda$ to account for the late-time accelerated expansion. We further assume an instantaneous reheating by the decay of $C$ for simplicity, and that the values of $C$, $\rho_{\rm rad}$ and $\rho_{\rm mat}$ shift before and after reheating as
\begin{eqnarray}
t_{\rm in} < t < t_{\rm reh} : && \!\!\!\!\!
\rho_{\rm mat} = \rho_{\rm rad} = 0 \, , \\
t_{\rm reh} < t : && \!\!\!\!\!
\frac{C}{a^3} + \rho_{\rm mat} = \frac{a_0^3}{a^3} \rho_{\rm mat}^0 \; , \;
\rho_{\rm rad} = \frac{a_0^4}{a^4} \rho_{\rm rad}^0 \, , \nonumber\\
\end{eqnarray}
where ``in'', ``reh'' and ``$0$'' denote the values at the instanton transition, reheating and present time, respectively.

The universe undergoes the standard cosmic history of the hot big bang cosmology after the reheating, namely nucleosynthesis followed by radiation-, matter- and then $\Lambda$-dominated periods. The fractional curvature ``density,'' defined as $\Omega_K(t) \equiv (3 K / a^2) / \rho_{\rm tot}$, evolves as
\begin{equation}
\frac{\vert \Omega_K (t_{\rm in}) \vert}{\vert \Omega_K (t_0) \vert} = 
\frac{\rho_0}{\rho_{\Lambda m}} 
\left( \frac{a_{\rm reh}}{a_{0}} \right)
\left( \frac{a_{0}}{a_{\Lambda m}} \right)^4
\left( \frac{a_{\Lambda m}}{a_{\rm eq}} \right) 
\left( \frac{a_{\rm in}}{a_0} \right) \; ,
\label{OmegaK_1}
\end{equation}
where the subscript ``eq'' and ``$\Lambda m$'' denote the values at the time of matter-radiation and $\Lambda$-matter equalities, respectively. The values of $\rho_0 / \rho_{\Lambda m}$, $a_{\rm reh} / a_0$, $a_0 / a_{\Lambda m}$ and $a_{\Lambda m} / a_{\rm eq}$ are given in the same way as the standard cosmological evolution. On the other hand, by setting $K = 1$, \eqref{eqn:ac3_over_tauc} gives
\begin{equation}
\frac{a_{\rm in}}{a_0} = \frac{s^{1/2}}{a_0} \left( \frac{\tau_{\rm in}}{a_{\rm in}} \right)^{1/2}
= \sqrt{\frac{s}{3} \, \vert \Omega_K (t_0) \vert \, \frac{\rho_0}{M_{\rm Pl}^2}}
\left( \frac{\tau_{\rm in}}{a_{\rm in}} \right)^{1/2} \; ,
\end{equation}
where $s \equiv \sqrt{\frac{3 \alpha_3}{2 ( 3 \lambda - 1 )}}$. Similarly, the fractional density $\Omega_K$ at the time $t = t_{\rm in}$ is approximately given by
\begin{equation}
\vert \Omega_K (t_{\rm in}) \vert \simeq \frac{3 / a_{\rm in}^2}{\alpha_3 / a_{\rm in}^6} =
\frac{3 s^2}{\alpha_3} \left( \frac{\tau_{\rm in}}{a_{\rm in}} \right)^2 \; .
\end{equation}
Hence \eqref{OmegaK_1} reduces to
\begin{eqnarray}
\frac{\tau_{\rm in}}{a_{\rm in}} & \simeq & \frac{\alpha_3^{2/3} \vert \Omega_K (t_0) \vert \, \rho_{\rm eq}^{2/9} \, \rho_0^{1/9}}{2^{8/9} \, 3 \, s \, \Omega_{m}^{8/9} \left( 1 + z_{\rm reh} \right)^{2/3}M_{\rm Pl}^{2/3}} 
\nonumber\\
& \simeq & 1.80 \cdot 10^{-47} \left(\frac{M_{\rm Pl}}{M}\right)^{2/3}
\left( \frac{3\lambda - 1}{2} \right)^{1/2} 
\nonumber\\ && \qquad  \times 
\left( \frac{10^{10}}{1 + z_{\rm reh}} \right)^{2/3} \, \frac{\vert \Omega_K(t_0) \vert}{0.005}
\; ,
\label{taucac-res}
\end{eqnarray}
where $M\equiv \alpha_3^{-1/4}$ as defined in (\ref{eqn:anisotropic-instanton-scaling}), and in the last approximate equality we have used the observed values to plug in $\rho_{\rm eq} \simeq \left( 5.67 \cdot 10^{-28} M_{\rm Pl} \right)^4$, $\rho_0 \simeq \left( 1.01 \cdot 10^{-30} M_{\rm Pl} \right)^4$ and $\Omega_{m} \simeq 0.308$. Since we have set $K=1$, the scale factor $a$ has the dimension of length. 

As we learn from \eqref{taucac-res}, we need an anisotropic instanton with the level of anisotropy of order $T/L \lesssim 10^{-47}$ in order to respect the observational upper bound $\vert \Omega_K(t_0) \vert \lesssim 0.005$ \cite{Ade:2015xua}, provided that the reheating occurs before BBN ($z_{\rm reh} \gtrsim 10^{10}$), that $\lambda$ at the time of tunneling does not deviate much from its (expected) IR value $\lambda_{\rm IR} = 1$, and that $M/M_{\rm Pl} \sim {\cal O}(1)$.
This small value is to account for the present flatness of the universe by the proposed mechanism in Sec.~\ref{sec:noboundary}, which in the inflationary cosmology would be compensated by the duration of inflation $\sim 50 - 60$ e-foldings. This also sets the lower bound on the energy scale that the instanton tunneling has to occur. At the time of this transition, \eqref{eqn:FriedmannEQ} gives 
\begin{eqnarray}
&&\frac{3(3\lambda-1)}{2}H_{\rm in}^2 \sim \frac{C}{a_{\rm in}^3} \sim
\frac{\alpha_3}{a_{\rm in}^6} \simeq
\frac{\alpha_3}{s^3} \left( \frac{a_{\rm in}}{\tau_{\rm in}} \right)^3 
\nonumber\\
&& \quad \simeq 
\left( 1.28 \cdot 10^{35} \right)^4 \frac{M^4}{M_{\rm Pl}^2}
\left( \frac{1 + z_{\rm reh}}{10^{10}} \right)^{2}
\left( \frac{0.005}{\vert \Omega_K(t_0) \vert} \right)^3 \, , 
\nonumber\\
\label{Hin2-concrete}
\end{eqnarray}
where $H_{\rm in}$ is the value of Hubble parameter at the time of instanton transition (in Planck units), and this corresponds to the energy scale at the transition to be $E_{\rm in} \equiv \sqrt{M_{\rm Pl}H_{\rm in}} \gtrsim 10^{35} M$. If we set $M=M_{\rm Pl}$ and if reheating occurs at the time $K/a^2 \sim (1/M^4)(K/a^2)^3$, i.e.~$\rho \sim C \sim 10^{56} z_{\rm reh} \vert \Omega_{K}(t_0) \vert^{-3/2}$, then one finds $z_{\rm reh} \sim 10^{60} \vert \Omega_K(t_0) \vert^{-1/2}$ and therefore from \eqref{Hin2-concrete}, $H_{\rm in}^2 \sim (10^{58})^4 \vert \Omega_K (t_0) \vert^{-7/2}$, which is consistent with the general results in Sec.~\ref{general}

\section{A more general solution to the flatness problem}
\label{app:moregeneral}

In Section~\ref{general} we showed how the concrete model presented in this paper may be part of a more general class of solutions. In this Appendix we expand further on this argument, both in scope and in terms of interpretation. 

There is a simple interpretation of the general argument presented in Section~\ref{general}. It is known that modified dispersion relations (MDR) may lead to an energy dependent speed of propagation for massless particles. This falls under the general umbrella term of ``varying speed of light'' (see~\cite{Magueijo:2003gj} for an early review). In the guise of MDRs, such theories lead to several astrophysical and cosmological implications (e.g.~\cite{AmelinoCamelia:2000mn,Amelino-Camelia:2013tla}).
The phenomenon may be quantified by the phase speed $c_p=E/p$ or the group speed $c_g=dE/dp$. 
In the case of  (\ref{MDR}) and  (\ref{MDR-ap}), in the UV we have:
\begin{equation}
c_p\propto c_g\propto \left(\frac{p}{M}\right)^{z-1} .
\end{equation}
In view of this, it is tempting to map the Friedmann equation (\ref{fried}) into the standard-looking Friedmann equation:
\begin{equation}\label{fried1}
H^2=\frac{1}{3}\rho -\frac{Kc_h^2}{a^2} 
\end{equation}
also with a time dependent $c$, and where we have reinstated $K$ as the culprit for the sign
ambiguity of the curvature term (relevant in what follows). Assuming $K\neq 0$ we have 
$c^2_h=a^2M^2 f$, so that in the UV:
\begin{equation}
c_h\approx 
\left(\frac{1}{Ma}\right)^{z-1}.
\end{equation}
We see that in the deep UV we have $c_h\propto c_g\propto c_p$,
(with the understanding that comparisons assume the replacements 
$E^2\rightarrow H^2$ and $p^2\rightarrow |K|/a^2$). Thus in the deep UV the various $c$ may be used interchangeably. The transition from UV to IR may be different, but this is not important here. 
 
This interpretation at once connects the solution of the flatness problem presented here to that in~\cite{Albrecht:1998ir}. This is particularly relevant if we wish to consider the implication of non-conservation of energy mentioned above.
As shown in~\cite{Albrecht:1998ir}  such violations actually help solving the flatness problem, reinforcing the argument. 

As is well known, violations of Lorentz invariance may bring about non-conservation. This depends on how we close the system started by (\ref{fried1}). In the concrete model presented in this paper, conservation of $\rho$ is assumed (or rather, one starts from the second Friedmann equation and then integrates it into the first, building conservation into the model). An alternative is to assume no modifications to the second Friedmann equation:
\begin{equation}\label{fried2}
\frac{\ddot a}{a }= -\frac{1}{6} \rho (1+3w).
\end{equation}
This implies violations of the Bianchi identities and energy conservation. Specifically, in combination with (\ref{fried1}) we find:
\begin{equation}
\dot \rho +3\frac{\dot a}{a}\rho(1+w)=\frac{6Kc^2}{a^2}\frac{\dot c}{c}.
\end{equation}
Merely looking at the sign of the RHS is very informative.
Defining $\rho_{\rm in}=3H^2$ we see at once that if $\dot c/c<0$ the violations of energy conservation act so as to push the Universe towards flatness. If the universe is closed ($K=1$ and thus supercritical, $\rho>\rho_{\rm in}$) then energy is removed from the universe; if the universe is open ($K=-1$ and $\rho<\rho_{\rm in}$) then energy is inserted into the universe; no violations occur for a flat model. Thus $\rho$ is pushed to $\rho_{\rm in}$. 

This does not mean that these violations are needed, or indeed relevant in all regimes. As in~\cite{Albrecht:1998ir} we can combine (\ref{fried1}) and (\ref{fried2}) to obtain:
\begin{equation}\label{Omegaeq}
\dot\Omega_K=(1-\Omega_K)\Omega_K\frac{\dot a}{a}(1+3w)+ 2\frac{\dot{c}}{c}\Omega_K.
\end{equation}
If $\Omega_K\ll 1$ this integrates to:
\begin{equation}\label{flat-sol1}
\Omega_K\propto a^{1+3w} c^2
\end{equation} 
leading to (\ref{flat-sol}), which was obtained by ignoring violations of energy conservation. Thus these violations are not very important in the solution to the flatness problem, as long as curvature is already sufficiently suppressed. 

Where these violations may be interesting is in situations in which the universe does not start from exact equipartition. Let us consider an extreme case. Suppose that initially $\rho=0$ and $K=-1$, that is a Milne Universe beginning. Then the Universe starts with $\Omega_K=1$ and no matter. This would be hopeless if energy were conserved (the Universe would simply remain empty). However inserting this condition into (\ref{Omegaeq}), we see that the first term initially vanishes, but the second term leads to $\Omega_K\propto c^2$. Hence curvature is still suppressed (at this rate) while matter is being dumped into the Universe. $\rho=0$ is also pushed to $\rho=\rho_{\rm in}$. Eventually  $\Omega_K\ll 1$, after which violations of energy conservation become irrelevant, and suppression of curvature proceeds according to (\ref{flat-sol1}).

\end{document}